\documentclass[aps,prl,showpacs,showkeys,superscriptaddress,twocolumn,groupedaddress]{revtex4-2}
\usepackage{CJK}
\usepackage[colorlinks=true,linkcolor=blue,anchorcolor=blue,citecolor=blue,urlcolor=blue,breaklinks=true]{hyperref}
\usepackage{graphicx}% Include figure files
\usepackage{subfigure}
\usepackage{amsmath}
\usepackage{amssymb}
\usepackage{slashed}
\usepackage{latexsym}
\usepackage{epsfig}
\usepackage{amsbsy}
\usepackage{array}
\usepackage{bm}
\usepackage{lipsum}
\usepackage{mathrsfs}
\usepackage{float}
\usepackage{color}
\usepackage[T1]{fontenc}
\usepackage{mathptmx}
\usepackage{breakurl}
\DeclareMathAlphabet{\mathcal}{OMS}{cmsy}{m}{n}
\DeclareSymbolFont{largesymbols}{OMX}{cmex}{m}{n}
\newcommand{\ie}{\emph{i.e.}, }

\begin{document}
\begin{CJK*}{UTF8}{gbsn}
\author{Yonghui Xia (夏永辉)}
\email{xiayh@xiahuhome.com}
\affiliation{College of Engineering and Applied Sciences, Nanjing University, Nanjing 210093, China}

\author{Hongtao Feng (冯红涛)}
\email{fenght@seu.edu.cn}
\affiliation{School of physics, Southeast University, Nanjing 211189, China}

\date{\today}

\title{Chaos, the Critical Phenomenon in Phase Space: Feigenbaum Constants and Critical Exponents}
\begin{abstract}
\bigskip

Chaos in both dissipative systems and conservative systems is investigated on  the   approach of renormalization group. It is found that the chaos is regarded as the critical phenomenon of equilibrium statistics in phase space. The two Feigenbaum constants in the period-doubling bifurcation  systems correspond to two independent  critical exponents, which are universal and can be adopted to distinguish the classes of chaos. For the conservative systems, due to the critical nature of the chaos, the isolated systems with different parameters are correlated   in the phase space, and therefore the isolated system is no longer isolated in the phase space.  The information of conservative systems is irreversibly lost over time, which leads to the increase entropy  in  an isolated  system, and the  contradiction between the second law of thermodynamics and the reversibility of isolated systems can be resolved. 
\end{abstract}
\maketitle
\end{CJK*}
\section{Introduction}
 Chaos, which is given by the deterministic  systems, widely exists in different fields, for example, physics, biology, economics,   atmospheric science,  and so on~\cite{Tabor:1989, Rasband:2015}.  It seems to be \emph{random} and \emph{disordered}, and therefore  it  is not  predictable, \emph{e.g.}, a typical chaotic mapping,  horseshoe map, is one-one corresponding to a sequence of  coin-flipping experiment~\cite{Smale:1967, Smale:1998}.  However, it still shows some  regular patterns, \ie the self-similarity structure,  in phase space~\cite{Lorenz:1963}.  Also,  the universal  Feigenbaum constants in period-doubling bifurcation  systems are found on the route to chaos~\cite{Feigenbaum:1978, Feigenbaum:1979, Lanford:1982, Lyubich:1999}.   The nature of chaos, which possesses both chaotic  and regular behaviors, is fascinating. In this letter, we find that, \emph{chaos, both in dissipative systems and conservative systems,  is the critical phenomenon in phase space}.      For the  chaos given by the period-doubling bifurcation systems, the two Feigenbaum constants are corresponding to the independent critical exponents of phase transition in phase space respectively. 

\begin{figure}[htp!]
\centering
\subfigure[]{\includegraphics[scale=0.30]{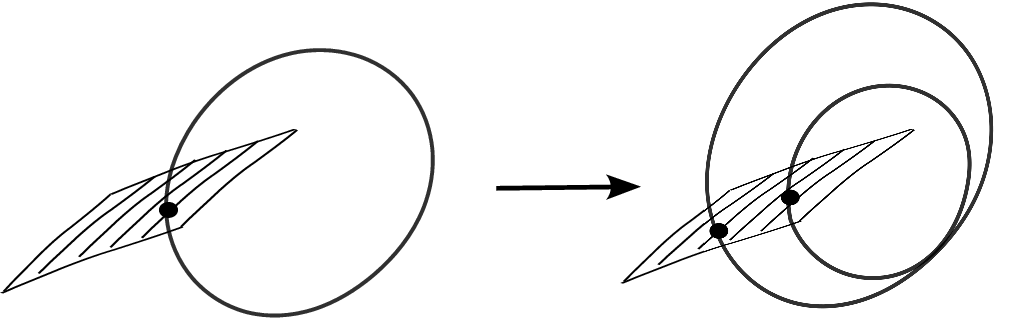}\label{Limit_cycle}}
\subfigure[]{\includegraphics[scale=0.60]{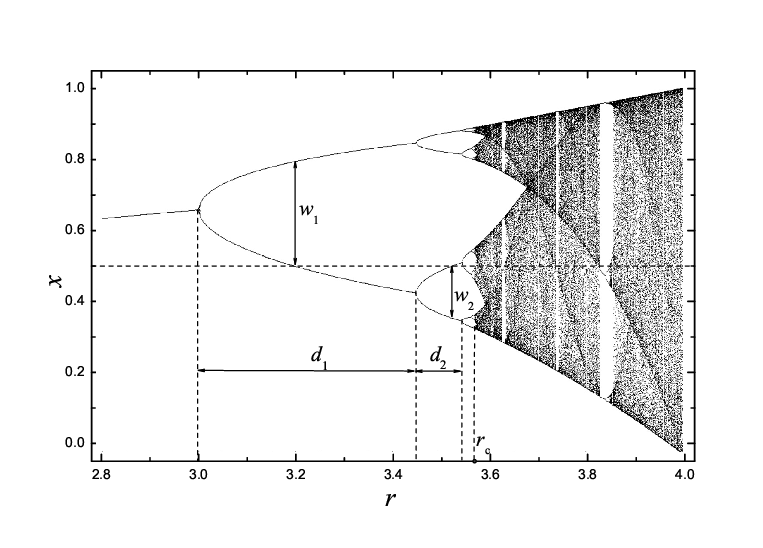}\label{bifurcation_diagram}}
\caption{(a). The sketch of the period--doubling process in the continuous dynamical system,  period--$2^{n}$ cycle $\rightarrow$ period--$2^{n+1}$ cycle. (b). The bifurcation diagram given by the logistic map $x_{i+1}=rx_{i}(1-x_{i}) $.  }
\label{bifurcation_diagram-Limit_cycle}
\end{figure}

\section{Chaos as the critical phenomenon in phase space}
\subsection{The chaos in dissipative systems}
  Let us  start from the period-doubling bifurcation  systems---- the typical dissipative dynamical systems. 
 Considering  the universality of the period-doubling bifurcation  systems~\cite{Feigenbaum:1978, Feigenbaum:1979, Lanford:1982, Lyubich:1999},  the simple  iterated function, named  logistic map~\cite{May:1976}, $x_{i+1}=rx_{i}(1-x_{i})$,  is taken as an example. This discrete dynamical system can be given by a  continuous  dynamical system through the Poincar\'e map, shown in  Fig.~\ref{Limit_cycle}.  From Fig.~\ref{bifurcation_diagram},  the two Feigenbaum constants  are given by 
\begin{eqnarray*}   
\delta=\lim_{n\rightarrow\infty} \frac{d_{n}}{d_{n+1}}=4.6692\cdots\; \text{and}\; \alpha=\lim_{n\rightarrow\infty}\frac{ w_{n}}{w_{n+1}}=2.5029\cdots
\end{eqnarray*}
 respectively, where the distances between the neighbored   bifurcation points are denoted by $d_n$, and the distances  of the nearest   neighbored   cycle elements in the $2^n$ cycle to $x=0.5$ are denoted by $w_n$. The universality of these two constants is confirmed by a rigorous mathematical proof~\cite{Lyubich:1999},  and they are found in different systems~\cite{Maurer:1979,*Libchaber:1980,*Giglio:1981,*Linsay:1981,Simoyi:1982}.  

Indeed, as shown in Fig.~\ref{bifurcation_diagram}, across the critical value $r_c$,  the trajectories  are  self-similar with the chaotic parameters in the chaotic regime $[ r_c, 4.0]$, \ie fractal structures.   Moreover, the chaotic parameters in this regime form a fat cantor set, \ie a fat fractal, which labels the accessible space~\cite{Jakobson:1981, Farmer:1985,*Farmer:1986}, and the chaotic bands are regarded as the ``mass'' on this fat cantor set, then the trajectories in this interval  $[ r_c, 4.0]$  form a multifractal. Noticed here that the map in each chaotic band has absolutely continuous invariant measure~\cite{Lyubich:2002}.    According to ergodic theorem~\cite{Birkhoff:1931}, the statistical system given by the dynamical systems with the parameters in this regime is ergodic. Thus,  the time average of  the observable $\overline{\boldsymbol{O}}_{\tau}$ over a long  time  is exactly the expected value  of the  statistical average $\overline {\boldsymbol{O}}$, $\overline {\boldsymbol{O}}_{\tau\rightarrow \infty}=\overline{\boldsymbol{O}}$.  Then,  this system can be regarded as a two-dimensional statistical system in equilibrium.  Actually, this  multifractal nature means that the system is at the \emph{critical point} in the parameter space, which is similar to  the critical point of continuous phase transition and is shown clearly in the following renormalization group calcualtion. 
 In this sense, the deviation from initial conditions, $\Delta$, can be regarded as the fluctuation in  phase space, $\Delta=\sqrt{\overline{(\boldsymbol{O}-\overline{\boldsymbol{O}})^2}}$.  Ones can easily find that the ``static susceptibility'', which is the  linear response to an external field,  at~(near) the critical point,
\begin{eqnarray}  
\chi\propto\mathcal{G}({p_\zeta}=0)
&=&\left. \int   \mathrm{d}{\zeta}\exp(-ip_{\zeta}\cdot\zeta)\mathcal{G}({\zeta},0)\right|_{{p_\zeta}=0} \nonumber\\
&=&\int   \mathrm{d}{\zeta}\overline{\left(\boldsymbol{O}({\zeta})- \overline{\boldsymbol{O}({\zeta})}\right)\; \left( \boldsymbol{O}(0)-\overline{\boldsymbol{O}(0)}\right)}\nonumber\\ \label{static susceptibility}
\end{eqnarray}
 is divergent, $\chi\rightarrow\infty$. 
Here, the ``distance'' ${\zeta}$ denotes the distance between two points in  phase space. As shown in Eq.~(\ref{static susceptibility}), the divergent ``susceptibility''  means  that the correlation length is infinite,   and  also one cannot obtain the information of state of the system from the ``initial fluctuation''.  Specifically,  since the fluctuations between different scales are strong at the critical point,   all the points in the interval $[r_c,4]$ are correlated, then one cannot obtain the information of the state of the system unless all the correlation  information between all  the points  in   phase space is known.  However, it is impossible, even in theory.  Due to that only the evaluation equation of the dynamical system  is known,  the lost correlation information cannot be obtained  even all the trajectories in the chaotic parameter interval are considered, \ie the information is permanently lost. Therefore, the unknown information, \ie ``lack of information'', makes the chaotic system  ``\emph{random}'',   and then one cannot predict the far future with a tiny deviation from initial conditions.  In this sense, the ``\emph{probability}'' is a suitable language  to describe  the ``\emph{random}'' nature of chaos.

According to the above arguments,  the parameter space where this critical point  lives,  
should be defined. As is known, the  periods of the cycles in the regimes between  two neighbored  bifurcation points  keep the same.  Actually,  these cycles in the regime with the same period  can be regarded as a statistical system and a set of parameters can be adopted to describe its statistical nature, which are similar to the roles of temperature and density.   
In this sense, $d$ and $w$, which label the regime,   play the role of ``temperature'' and ``density''. 
Then, the chaotic state  actually refers to a state at the critical point in  the parameter  space $d-w$, and the bifurcation process can be considered as the process of adjusting these two  parameters to make the system approach the critical point. 

Then, let us  investigate the correlation length  in the period-doubling bifurcation process.  Here, a three--dimensional  continuous dynamical system corresponding to Fig~\ref{bifurcation_diagram} is taken into consideration.   
The cycles in phase space can be described by a   periodic function, for example, the  periodic function $\mathcal{F}_{\boldsymbol{K}_{\zeta}}(\zeta+\boldsymbol{a})=\mathcal{F}_{\boldsymbol{K}_{\zeta}}(\zeta)$ can describe these period-$2^n$ cycles. According to Bloch's theorem~\cite{Bloch:1929}, this periodic function can be given by the solution of   periodic linear differential function, for example, Schr\"odinger equation in  a periodic potential, which is shown clearly for   the case of crystals. Specifically, the solution of this three-dimensional  linear differential function~(Schr\"odinger equation) in  phase space is given by  $\mathcal{F}(\zeta^{\prime})=\mathcal{F}(\zeta+\boldsymbol{a})=\exp(i\boldsymbol{K}_{\zeta}\boldsymbol{a})\mathcal{F}(\zeta)$,  in which  $\mathcal{F}(\zeta)=\exp(i\boldsymbol{K}_{\zeta}{\zeta})\mathcal{F}_{\boldsymbol{K}_{\zeta}}(\zeta)$,  $K_{\zeta i}\in[-\pi/\boldsymbol{a}_i, \pi/\boldsymbol{a}_i], (i=(1,2,3)$, $\boldsymbol{K}_{\zeta}=(\boldsymbol{K}_{\zeta 1},\boldsymbol{K}_{\zeta 2},\boldsymbol{K}_{\zeta 3})$,   and $\boldsymbol{a}=(\boldsymbol{a}_1,\boldsymbol{a}_2,\boldsymbol{a}_3))$. 
Here, $\mathcal{F}_{\boldsymbol{K}_{\zeta}}(\zeta)$  can be regarded as the ``Bloch wave function'', and therefore,  $\boldsymbol{a}$ is the ``lattice constant'' in the phase space.  
 According to  Nambu-Goldstone's  theorem~\cite{Nambu:1960,*Goldstone:1961,*Goldstone:1962}, because of the relation  $\mathcal{F}(\zeta) \xrightarrow{\boldsymbol{K}_{\zeta}=0}\mathcal{F}_{\boldsymbol{K}_{\zeta}}(\zeta)$, $\mathcal{F}_{\boldsymbol{K}_{\zeta}}(\zeta)$ can be regarded as the description of the symmetry broken phase on the ``lattice'', \ie ordered phase. In this sense,   the system is ordered, and the dynamical system  is actually in   a self--organized state~\cite{Nicolis:1978}. Also, the trajectories of the corresponding discrete dynamical system given by these cycles can be  regarded as  a symmetry broken phase on the ``lattice''. Noticed here that, the breaking symmetry is not a continuous symmetry in the discrete dynamical system.  For the regime with period-$2^{n+1}$ cycles, due to the doubling of the period, the ``lattice constant''  is  half of   that  with  period-$2^{n}$ cycles. Then, according to   the renormalization approach in the case of critical behavior~\cite{Wilson:1971A,*Wilson:1971B},  one has the  zoom multiple $L$ in the bifurcation process, 
\begin{equation}
{L}=\frac{\boldsymbol{a}}{\boldsymbol{a}/2}=2,
\end{equation}
and  the corresponding correlation lengths with the parameters $d_{n+1}$ and $w_{n+1}$  are  therefore  doubled respectively,
\begin{eqnarray}
\frac{\xi(d_n)}{\xi(d_{n+1})}=\frac{\xi(w_n)}{\xi(w_{n+1})}=\frac{1}{L}=\frac{1}{2}.
\end{eqnarray} 
This relation shows clearly that this multifractal  corresponds to an unstable fixed point, the critical point.

For $n \rightarrow \infty$,   $d_{n\rightarrow\infty}$ and $w_{n\rightarrow\infty}$ are both zero. Thus, the system with the parameters in  the interval $ [r_c, 4]$, as shown in Fig.~\ref{bifurcation_diagram},  is at the critical point with the parameter set $(d=0,w=0)$. 
For  the parameters $(d,w)$ near the critical point~$(0,0)$,  the critical exponents can be computed in the following formulas, 
\begin{eqnarray}
\frac{\xi(d_n)}{\xi(d_{n+1})}=\left( \frac{|d_{n}-0|}{|d_{n+1}-0|}\right)^{-\nu_{1}}, 
\frac{\xi(w_n)}{\xi(w_{n+1})}=\left( \frac{|w_{n}-0|}{|w_{n+1}-0|}\right)^{-\nu_{2}},\nonumber\\
\end{eqnarray}
where $\nu_{1}$ and $\nu_{2}$ are the critical exponents.
As the bifurcation number approaches infinity,  we have~\cite{Feigenbaum:1978,Lyubich:1999}
\begin{eqnarray}
\lim_{n\rightarrow\infty} \frac{|d_{n}-0|}{|d_{n+1}-0|}=\delta,\;\;\lim_{n\rightarrow\infty} \frac{|w_{n}-0|}{|w_{n+1}-0|}=\alpha,
\end{eqnarray}
and therefore the relations between the critical exponents and the two Feigenbaum constants are respectively given by
\begin{eqnarray*}
\nu_{1} =\log_{\delta} {2} = 0.44981\cdots,
\nu_{2} =\log_{\alpha}{2}= 0.75551\cdots.
\end{eqnarray*}
As is known, the critical exponents  are universal, which are independent of the details of the local interaction,  and they can be adopted to distinguish the classes of  phase transitions. Thus, the Feigenbaum constants are also independent of the details of the dynamical system equations~\cite{Feigenbaum:1978, Feigenbaum:1979, Lanford:1982, Lyubich:1999}, and they can label the classification of chaos.  In other words, the dynamical systems with the same Feigenbaum constants are in a same class.

 It is known that, the system in chaotic state can be predicted in a short time, which corresponds to the short--range trajectory in phase space. With the increase of time, the long--range correlation is built from the short--range trajectories, \ie the self-similar pattern is built with time. 
 In this sense, for a long time, the   trajectory with a fixed parameter $r\;(r\geq r_c)$ in chaotic regimes, which is  determined by a determined evolution equation, is correlated with all the trajectories with  the parameters in the  interval $[r_c, 4]$ respectively. However, as mentioned above, the correlation information cannot be given by  the determined evolution equation with a fixed parameter,  and thus the correlation information between the trajectories with different parameters is lost over time.
 Then the long time behavior cannot be predicted: ones cannot know all the correlation information from the  evolution equation with a fixed parameter $r$,  and thus the Kolmogorov-Sinai entropy~\cite{Shiryayev:1993, *Kolmogorov:1959, Sinai:2010} is nonzero in these regimes, and then the system shows the random behavior.
 
 According to the above arguments, chaos can be investigated as a phase transition problem in phase space.   
 Noticed that this  phase transition is in  an \emph{equilibrium statistical system},
  although the  period-doubling bifurcation  system is a dissipative system, which is considered to be far away from equilibrium.  
  For a continuous dynamical system, as the discussions above, the symmetry of the corresponding statistical system  should be spontaneously broken in the Nambu-Goldstone mode.
 Because of the Mermin--Wagner theorem~\cite{Mermin:1966,*Coleman:1973}, the phase transition  due to the spontaneous breaking of continuous global symmetry cannot undergo in  less than three-dimensional systems. Although a \emph{Berezinskii--Kosterlitz--Thouless}~({BKT}) phase transition  exists in the two-dimensional systems, whose correlation length is also divergent at the critical point, it actually describes the phase transition between the quasi-long-range ordered phase and disordered phase~\cite{Berezinskii:1971,*Berezinskii:1972, Kosterlitz:1973}. In this sense, chaos cannot exist in one or two- dimensional continuous  nonlinear dynamical systems, which  agrees with Poincar{\'e}--Bendixson theorem~\cite{Poincare:1881, *Bendixson:1901}. 
 For the discrete dynamical systems, the Mermin--Wagner theorem does not work, the chaos can exist.
  Besides, it is known that the chaotic behavior can also be found in an infinite dimensional linear system~\cite{Bonet:2001}, and this fact agrees with  the case of phase transition in a quantum system, which is linear and lives in an infinite dimensional  Hilbert space.

\subsection{ The chaos in conservative systems}
  For the  conservative systems, the conclusions given by the dissipative systems still hold. For example, the minimal freedom of degrees of the system, where the minimal stochastic webs~(chaos) exist, is  $N_c=3/2$~\cite{Zaslavsky:1986, *Zaslavsky:1988,Chernikov:1987a, *Chernikov:1988,*Chernikov:1987b},  which corresponds to  the minimal dimensions of  spontaneous   continuous global symmetry breaking, \ie three dimensions~($2N_c=3$). Indeed, similar to the dissipative systems,     the  trajectories of the conservative systems  in the chaotic state are also self-similar in the phase space. Moreover, the self-similarity of such conservative systems  manifests itself as fat fractals~\cite{Mandelbrot:1983, Mayer-Kress:1986,Pietronero:1986,Umberger:1985}. Noticed here that for some strongly chaotic systems, for example, the  Sinai billiard system, which is ergodic and its trajectory in phase space is not regarded as a fractal, however, considering its self-similar structure in the phase space and nonzero measure of chaos~\cite{chernov:2006}, we can also regard the trajectory as a ``fat fractal''. Actually, due to the nonzero Lebesgue measure~\cite{Meiss:1992}, the trajectory always forms a fat fractal in the chaotic state as the system crosses the  critical parameter, and the phase space  given by this parameters interval cannot be decomposed into independent parts.   In this sense, similar to the dissipative systems, in this parameters interval, all the trajectories with different parameters form a multifractal. This multifractal nature means that the conservative dynamical systems with energy conservation, which are given by the different parameters in this interval,  are not independent, although there do not exist exchanges of energy and matter between them. Moreover,  according to Liouville's theorem~\cite{Gibbs:1902}, the evolution of a conservative system is measure-preserving, and therefore this  statistical system given by these dynamical systems in  this parameter interval is ergodic~\cite{Birkhoff:1931}. Thus, chaos in conservative systems  can also be regarded as  a critical phenomenon in phase space.  The above arguments show that there does not exist anomalous dimension in  this  multifractal.  
  Indeed, this  multifractal corresponds to the critical point, where   the mean field approximation works well~\cite{Landau:2013}.

 Similar to dissipative dynamical  systems, for the conservative systems, the long-range correlation, \ie the self-similar structure,  builds up gradually over time at the chaotic state, and then the information of correlation  between  the trajectories with  different parameters is also lost over time, \ie the entropy increases with time. In this sense, the nonlinear (chaotic) map in the chaotic state is not a bijection  but  an  injection, and thus  the dynamical system is irreversible.  In fact, the contradiction between the second law of thermodynamics and  the Poincar\'{e} recurrence theorem for the statistical system~\cite{Poincare:1890,Caratheodory:1919} does not exist: though the  trajectory is  arbitrarily close to the initial state  for an enough long time, due to the lost information, the corresponding states are different. 

  Actually, this picture of the lost information  is similar to the case of decoherence process in  quantum system~\cite{Zeh:1970,Zurek:1981, *Zurek:1982}.   Noted  that, since the measured object and the environment together  evolve according to the deterministic Schr\"{o}dinger equation, the decoherence theory cannot solve the defined outcome problem in quantum mechanics, which has to resort to the quantum mechanics interpretations. However,  the conservative system evolves over time according to the deterministic  dynamical system equation(s), and one therefore has a defined outcome. As we mentioned above, the information of correlation is lost permanently.  This fact implies that the ``real'' randomness of quantum mechanics hides in classical mechanical systems, which are usually nonlinear. 
  In this sense,  we  can linearize the nonlinear dynamical system equations and the state can  be described by ``wave function''~(state vector) in the phase space. Considering the topological nontrivial nature of the fat fractal, for an isolated system, although the coordinate in the  phase space given by the final state   is arbitrarily  close to  that given by the  initial state in the chaotic state, the corresponding states of this dynamical system are not close.  
In detail, the stable islands exist in the chaos of conservative systems, which can be regarded as the holes in the phase space~\cite{Mandelbrot:1983, Mayer-Kress:1986,Pietronero:1986,Umberger:1985},  and they play the role of solenoid in \emph{Aharonov-Bohm} effect~\cite{Aharonov:1959}.

\section{Conclusions} 
To summarize, according to the renormalization  group approach of equilibrium statistics, the chaos is regarded as the critical phenomenon in phase space  for either  dissipative systems or conservative systems. It is found that the  two Feigenbaum constants in the   period-doubling bifurcation  systems are corresponding to two independent  critical exponents, which are universal, and thus they  can be adopted to distinguish the classes of chaos.  For the chaos in conservative systems, similar to the decoherence process in the quantum systems, the isolated systems with different parameters are correlated in the phase space, and the information of correlations is lost over time permanently, which is  irreversible. Thus, the  contradiction between the second law of thermodynamics and the reversibility of conservative systems can be resolved.

\section{Acknowledgment}
\begin{CJK*}{UTF8}{gbsn}
This work is supported by the National Natural Science Foundation of China (Grant No. 12233002).  Yonghui Xia  is  grateful to the supports from  his wife Ms. Q. Hu (胡倩) and his lovely angels  R.Q. Xia (夏若缺) and H.Z.Y. Xia (夏胡至一); Hongtao Feng thanks his children, N.H.  Feng (冯诺恒) and N.L. Hong (洪诺琳), for their encouragement.  
\end{CJK*} 
\bibliography{reference}

@Article{Feigenbaum:1978,
	author = {Feigenbaum, Mitchell J.},
	day = {01},
	doi = {10.1007/BF01020332},
	issn = {1572-9613},
	journal = {J. Stat. Phys.},
	month = {Jul},
	number = {1},
	pages = {25--52},
	title = {Quantitative universality for a class of nonlinear transformations},
	url = {https://doi.org/10.1007/BF01020332},
	volume = {19},
	year = {1978}}

@Article{Lorenz:1963,
	address = {Boston MA, USA},
	author = {Edward N. Lorenz},
	doi = {10.1175/1520-0469(1963)020<0130:DNF>2.0.CO;2},
	journal = {J. Atmos. Sci.},
	number = {2},
	pages = {130 - 141},
	publisher = {American Meteorological Society},
	title = {Deterministic Nonperiodic Flow},
	url = {https://journals.ametsoc.org/view/journals/atsc/20/2/1520-0469_1963_020_0130_dnf_2_0_co_2.xml},
	volume = {20},
	year = {1963}}

@book{Rasband:2015,
	author = {Rasband, S Neil},
	note = {and references therein},
	publisher = {Courier Dover Publications},
	title = {Chaotic dynamics of nonlinear systems},
	year = {2015}}

@Article{Wilson:1971A,
	author = {Wilson, Kenneth G.},
	doi = {10.1103/PhysRevB.4.3174},
	issue = {9},
	journal = {Phys. Rev. B},
	month = {Nov},
	numpages = {0},
	pages = {3174--3183},
	publisher = {American Physical Society},
	title = {Renormalization Group and Critical Phenomena. \text{I}. Renormalization Group and the Kadanoff Scaling Picture},
	url = {https://link.aps.org/doi/10.1103/PhysRevB.4.3174},
	volume = {4},
	year = {1971}}

@Article{Wilson:1971B,
	author = {Wilson, Kenneth G.},
	doi = {10.1103/PhysRevB.4.3184},
	issue = {9},
	journal = {Phys. Rev. B},
	month = {Nov},
	numpages = {0},
	pages = {3184--3205},
	publisher = {American Physical Society},
	title = {Renormalization Group and Critical Phenomena. \text{II}. Phase-Space Cell Analysis of Critical Behavior},
	url = {https://link.aps.org/doi/10.1103/PhysRevB.4.3184},
	volume = {4},
	year = {1971}}

@book{Tabor:1989,
	author = {Tabor, Michael},
	note = {and references therein},
	publisher = {Wiley-Interscience},
	title = {Chaos and integrability in nonlinear dynamics: an introduction},
	url = {https://www.wiley.com/en-us/Chaos+and+Integrability+in+Nonlinear+Dynamics%3A+An+Introduction-p-9780471827283},
	year = {1989}}

@Article{Lyubich:1999,
	author = {Mikhail Lyubich},
	issn = {0003486X},
	journal = {Ann. Math.},
	number = {2},
	pages = {319--420},
	publisher = {Annals of Mathematics},
	title = {\text{Feigenbaum-Coullet-Tresser} Universality and Milnor's Hairiness Conjecture},
	url = {http://www.jstor.org/stable/120968},
	urldate = {2022-11-03},
	volume = {149},
	year = {1999}}

@Article{Mermin:1966,
	author = {Mermin, N. D. and Wagner, H.},
	doi = {10.1103/PhysRevLett.17.1133},
	issue = {22},
	journal = {Phys. Rev. Lett.},
	month = {Nov},
	numpages = {0},
	pages = {1133--1136},
	publisher = {American Physical Society},
	title = {Absence of Ferromagnetism or Antiferromagnetism in One- or Two-Dimensional Isotropic Heisenberg Models},
	url = {https://link.aps.org/doi/10.1103/PhysRevLett.17.1133},
	volume = {17},
	year = {1966}}

@Article{Coleman:1973,
	author = {Coleman, Sidney},
	day = {01},
	doi = {10.1007/BF01646487},
	issn = {1432-0916},
	journal = {Commun. Math. Phys.},
	month = {Dec},
	number = {4},
	pages = {259--264},
	title = {There are no Goldstone bosons in two dimensions},
	url = {https://doi.org/10.1007/BF01646487},
	volume = {31},
	year = {1973}}

@Article{Berezinskii:1971,
	author = {Berezinskii, VL},
	journal = {Sov. Phys. JETP},
	number = {3},
	pages = {493--500},
	title = {Destruction of long-range order in one-dimensional and two-dimensional systems having a continuous symmetry group \text{I}. Classical systems},
	url = {http://www.jetp.ras.ru/cgi-bin/dn/e_032_03_0493.pdf},
	volume = {32},
	year = {1971}}

@Article{Berezinskii:1972,
	author = {Berezinskii, VL},
	journal = {Sov. Phys. JETP},
	number = {3},
	pages = {610--616},
	title = {Destruction of long-range order in one-dimensional and two-dimensional systems possessing a continuous symmetry group. \text{II}. Quantum systems},
	url = {http://www.jetp.ras.ru/cgi-bin/dn/e_034_03_0610.pdf},
	volume = {34},
	year = {1972}}

@Article{Kosterlitz:1973,
	author = {J M Kosterlitz and D J Thouless},
	doi = {10.1088/0022-3719/6/7/010},
	journal = {J. Phys. C: Solid State Phys.},
	month = {apr},
	number = {7},
	pages = {1181},
	title = {Ordering, metastability and phase transitions in two-dimensional systems},
	url = {https://dx.doi.org/10.1088/0022-3719/6/7/010},
	volume = {6},
	year = {1973}}

@Article{Lanford:1982,
	author = {Lanford. III, OSCAR E},
	journal = {Bull. Amer. Math. Soc.},
	number = {3},
	title = {A computer-assisted proof of the Feigenbaum conjectures},
	url = {https://www.ams.org/journals/bull/1982-06-03/S0273-0979-1982-15008-X/home.html},
	volume = {6},
	year = {1982}}

@Article{Feigenbaum:1979,
	author = {Feigenbaum, Mitchell J.},
	day = {01},
	doi = {10.1007/BF01107909},
	issn = {1572-9613},
	journal = {J. Stat. Phys.},
	month = {Dec},
	number = {6},
	pages = {669--706},
	title = {The universal metric properties of nonlinear transformations},
	url = {https://doi.org/10.1007/BF01107909},
	volume = {21},
	year = {1979}}

@Article{Bendixson:1901,
	author = {Bendixson, Ivar},
	journal = {Acta Math.},
	pages = {1--88},
	publisher = {Institut Mittag-Leffler},
	title = {Sur les courbes d{\'e}finies par des {\'e}quations diff{\'e}rentielles},
	url = {https://doi.org/10.1007/BF02403068},
	volume = {24},
	year = {1901}}

@Article{Poincare:1881,
	author = {H. Poincar{\'e}},
	journal = {Journal de Math{\'e}matiques Pures et Appliqu{\'e}es},
	pages = {375-422},
	title = {M{\'e}moire sur les courbes d{\'e}finies par une {\'e}quation diff{\'e}rentielle (\text{I})},
	url = {http://eudml.org/doc/235914},
	volume = {7},
	year = {1881}}

@Article{Bloch:1929,
	author = {Bloch, Felix},
	day = {01},
	doi = {10.1007/BF01339455},
	issn = {0044-3328},
	journal = {Z. Physik},
	month = {Jul},
	number = {7},
	pages = {555--600},
	title = {{\"U}ber die Quantenmechanik der Elektronen in Kristallgittern},
	url = {https://doi.org/10.1007/BF01339455},
	volume = {52},
	year = {1929}}

@Article{Birkhoff:1931,
	author = {George D. Birkhoff},
	doi = {10.1073/pnas.17.2.656},
	journal = {Proc. Natl. Acad. Sci.},
	number = {12},
	pages = {656-660},
	title = {Proof of the Ergodic Theorem},
	url = {https://www.pnas.org/doi/abs/10.1073/pnas.17.2.656},
	volume = {17},
	year = {1931}}

@book{Landau:2013,
	author = {Landau, L.D. and Lifshitz, E.M.},
	isbn = {9780080570464},
	number = {Volume 5},
	publisher = {Elsevier Science},
	title = {Statistical Physics},
	year = {2013}}

@inproceedings{Farmer:1986,
	address = {Berlin, Heidelberg},
	author = {Farmer, J. D.},
	booktitle = {Dimensions and Entropies in Chaotic Systems},
	editor = {Mayer-Kress, Gottfried},
	isbn = {978-3-642-71001-8},
	pages = {54--60},
	publisher = {Springer Berlin Heidelberg},
	title = {Scaling in Fat Fractals},
	year = {1986}}

@Article{Umberger:1985,
	author = {Umberger, David K. and Farmer, J. Doyne},
	doi = {10.1103/PhysRevLett.55.661},
	issue = {7},
	journal = {Phys. Rev. Lett.},
	month = {Aug},
	numpages = {0},
	pages = {661--664},
	publisher = {American Physical Society},
	title = {Fat Fractals on the Energy Surface},
	url = {https://link.aps.org/doi/10.1103/PhysRevLett.55.661},
	volume = {55},
	year = {1985}
}

@Article{Smale:1967,
  	title={Differentiable dynamical systems},
  	author={Smale, Stephen},
  	journal={Bull.  Amer. math. Soc.},
  	volume={73},
  	number={6},
  	pages={747--817},
  	year={1967},
  	url={https://www.ams.org/journals/bull/1967-73-06/S0002-9904-1967-11798-1/S0002-9904-1967-11798-1.pdf}
}

@Article{Smale:1998,
	author = {Smale, Steve},
	da = {1998/03/01},
	doi = {10.1007/BF03024399},
	id = {Smale1998},
	isbn = {1866-7414},
	journal = { Math. Intell.},
	number = {1},
	pages = {39--44},
	title = {Finding a horseshoe on the beaches of \text{Rio}},
	ty = {JOUR},
	url = {https://doi.org/10.1007/BF03024399},
	volume = {20},
	year = {1998}
}

@Article{Zeh:1970,
	author = {Zeh, H.  D. },
	da = {1970/03/01},
	doi = {10.1007/BF00708656},
	id = {Zeh1970},
	isbn = {1572-9516},
	journal = {Found. Phys.},
	number = {1},
	pages = {69--76},
	title = {On the interpretation of measurement in quantum theory},
	ty = {JOUR},
	url = {https://doi.org/10.1007/BF00708656},
	volume = {1},
	year = {1970}
}

@Article{Bonet:2001,
	author = {Bonet, Jos\'{e} and Mart\'{i}nez-Gim\'{e}nez, F\'{e} lix and Peris, Alfredo},
	title = {A Banach Space which Admits No Chaotic Operator},
	journal = {B. Lond. Math. Soc.},
	volume = {33},
	number = {2},
	pages = {196-198},
	doi = {https://doi.org/10.1112/blms/33.2.196},
	url = {https://londmathsoc.onlinelibrary.wiley.com/doi/abs/10.1112/blms/33.2.196},
	year = {2001}
}

@Article{May:1976,
	author="May, Robert M.",
	title="Simple mathematical models with very complicated dynamics",
	journal="Nature",
	year="1976",
	month="Jun",
	day="01",
	volume="261",
	number="5560",
	pages="459--467",
	issn="1476-4687",
	doi="10.1038/261459a0",
	url="https://doi.org/10.1038/261459a0"
}

@Article{Zurek:1981,
  	title = {Pointer basis of quantum apparatus: Into what mixture does the wave packet collapse?},
  	author = {Zurek, W. H.},
  	journal = {Phys. Rev. D},
  	volume = {24},
  	issue = {6},
  	pages = {1516--1525},
  	numpages = {0},
  	year = {1981},
  	month = {Sep},
  	publisher = {American Physical Society},
  	doi = {10.1103/PhysRevD.24.1516},
  	url = {https://link.aps.org/doi/10.1103/PhysRevD.24.1516}
}

@Article{Zurek:1982,
  	title = {Environment-induced superselection rules},
  	author = {Zurek, W. H.},
  	journal = {Phys. Rev. D},
  	volume = {26},
  	issue = {8},
  	pages = {1862--1880},
  	numpages = {0},
  	year = {1982},
  	month = {Oct},
  	publisher = {American Physical Society},
  	doi = {10.1103/PhysRevD.26.1862},
  	url = {https://link.aps.org/doi/10.1103/PhysRevD.26.1862}
}

@Article{Farmer:1985,
  	title = {Sensitive dependence on parameters in nonlinear dynamics},
  	author = {Farmer, J. Doyne},
  	journal = {Phys. Rev. Lett.},
  	volume = {55},
  	issue = {4},
  	pages = {351--354},
  	numpages = {0},
  	year = {1985},
  	month = {Jul},
  	publisher = {American Physical Society},
  	doi = {10.1103/PhysRevLett.55.351},
  	url = {https://link.aps.org/doi/10.1103/PhysRevLett.55.351}
}

@Article{Jakobson:1981,
	author = {M. V. Jakobson},
	title = {{Absolutely continuous invariant measures for one-parameter families of one-dimensional maps}},
	volume = {81},
	journal = {Commun. Math. Phys.},
	number = {1},
	publisher = {Springer},
	pages = {39 -- 88},
	year = {1981},
	url={https://doi.org/10.1007/BF01941800}
}

@Article{Poincare:1890, 
	author={Poincar\'e, Henri},
	title={{Sur le probl\`eme des trois corps et les \'equations de la dynamique}},
	journal={Acta math.},
	volume={13},
	year={1890},
	url ={https://projecteuclid.org/journals/acta-mathematica/volume-13/issue-1-2},
pages={1--270} }

@Article{Caratheodory:1919,
	author = {C. Carath\'{e}odory},
	journal = {Berl. Sitzungsber},
	pages = {580--584},
	title = {{\"U}ber den Wiederkehrsatz von {Poincar\'e}},
	year = {1919}
}

@Article{Nicolis:1978,
	author = {G. Nicolis and I. Prigogine},
	title = { Self-Organization in Nonequilibrium Systems: From Dissipative Structures to Order through Fluctuations},
	journal = {Ber. Bunsenges. Phys. Chem.},
	volume = {82},
	number = {6},
	pages = {672-672},
	doi = {https://doi.org/10.1002/bbpc.197800155},
	url = {https://onlinelibrary.wiley.com/doi/abs/10.1002/bbpc.197800155},
	year = {1978}
}

@Book{Mandelbrot:1983,
  	title={The Fractal Geometry of Nature},
  	author={Mandelbrot, B.B.},
  	isbn={9780716711865},
  	lccn={97133578},
  	series={Einaudi paperbacks},
  	year={1983},
  	publisher={Henry Holt and Company}
}

@book{Mayer-Kress:1986,
  	title={Dimensions and Entropies in Chaotic Systems},
  	author={Mayer-Kress, G.},
  	isbn={978-3-642-71003-2},
  	series={Springer Series in Synergetics},
  	year={1986},
  	publisher={Springer Berlin, Heidelberg},
  	url={https://doi.org/10.1007/978-3-642-71001-8}
}

@book{Pietronero:1986,
  	title={Fractals in Physics},
  	author={Pietronero, L. and Tosatte, E.},
  	isbn={978-0-444-86995-1},
  	year={1986},
  	publisher={Elsevier B.V.},
  	url={https://doi.org/10.1016/C2009-0-14331-3}
}

@Article{Zaslavsky:1986,
       	author = {{Zaslavsky}, G.~M. and {Zakharov}, M. Iu. and {Sagdeev}, R.~Z. and {Usikov}, D.~A. and {Chernikov}, A.~A.},
        title = "{Stochastic web and pArticle diffusion in a magnetic field}",
      	journal = {Sov. Phys. JETP},
        year = 1986,
        month = aug,
      	volume = {91},
        pages = {500-516},
       	url = {http://jetp.ras.ru/cgi-bin/dn/e_064_02_0294.pdf}
}

@Article{Chernikov:1987a,
	title = {Some peculiarities of stochastic layer and stochastic web formation},
	journal = {Phys. Lett. A},
	volume = {122},
	number = {1},
	pages = {39-46},
	year = {1987},
	issn = {0375-9601},
	doi = {https://doi.org/10.1016/0375-9601(87)90772-9},
	url = {https://www.sciencedirect.com/science/Article/pii/0375960187907729},
	author = {A.A. Chernikov and M.Ya. Natenzon and B.A. Petrovichev and R.Z. Sagdeev and G.M. Zaslavsky}
}

@Article{Chernikov:1988,
	title = {Strong changing of adiabatic invariants, KAM-tori and web-tori},
	journal = {Phys. Lett. A},
	volume = {129},
	number = {7},
	pages = {377-380},
	year = {1988},
	issn = {0375-9601},
	doi = {https://doi.org/10.1016/0375-9601(88)90006-0},
	url = {https://www.sciencedirect.com/science/Article/pii/0375960188900060},
	author = {A.A. Chernikov and M.Ya. Natenzon and B.A. Petrovichev and R.Z. Sagdeev and G.M. Zaslavsky}
}

@Article{Zaslavsky:1988,
	doi = {10.1070/PU1988v031n10ABEH005632},
	url = {https://dx.doi.org/10.1070/PU1988v031n10ABEH005632},
	year = {1988},
	month = {oct},
	publisher = {},
	volume = {31},
	number = {10},
	pages = {887},
	author = {G M Zaslavsky and  R Z Sagdeev and  D A Usikov and  A A Chernikov},
	title = {Minimal chaos, stochastic webs, and structures of quasicrystal symmetry},
	journal = {Usp. Fiz. Nauk}
}

@Article{Chernikov:1987b,
	author="Chernikov, A. A.
	and Sagdeev, R. Z.
	and Usikov, D. A.
	and Zakharov, M. Yu
	and Zaslavsky, G. M.",
	title="Minimal chaos and stochastic webs",
	journal="Nature",
	year="1987",
	month="Apr",
	day="01",
	volume="326",
	number="6113",
	pages="559--563",
	issn="1476-4687",
	doi="10.1038/326559a0",
	url="https://doi.org/10.1038/326559a0"
}

@Article{Maurer:1979,
  	title={Rayleigh-B{\'e}nard experiment in liquid helium; frequency locking and the onset of turbulence},
  	author={Maurer, J and Libchaber, A},
  	journal={J. Physique Lett.},
  	volume={40},
  	number={16},
  	pages={419--423},
  	year={1979},
  	publisher={Les Editions de Physique},
  	url={https://doi.org/10.1051/jphyslet:019790040016041900}
}

@Article{Libchaber:1980,
  	title={Une exp{\'e}rience de Rayleigh-B{\'e}nard de g{\'e}om{\'e}trie r{\'e}duite; multiplication, accrochage et d{\'e}multiplication de fr{\'e}quences},
  	author={Libchaber, Albert and Maurer, Jean},
 	journal={J. Physique Colloq.},
  	volume={41},
  	number={C3},
  	pages={C3--51},
  	year={1980},
  	publisher={EDP Sciences},
  	url={https://doi.org/10.1051/jphyscol:1980309}
}

@Article{Giglio:1981,
  	title = {Transition to Chaotic Behavior via a Reproducible Sequence of Period-Doubling Bifurcations},
  	author = {Giglio, Marzio and Musazzi, Sergio and Perini, Umberto},
  	journal = {Phys. Rev. Lett.},
  	volume = {47},
  	issue = {4},
  	pages = {243--246},
  	numpages = {0},
  	year = {1981},
  	month = {Jul},
  	publisher = {American Physical Society},
  	doi = {10.1103/PhysRevLett.47.243},
  	url = {https://link.aps.org/doi/10.1103/PhysRevLett.47.243}
}

@Article{Linsay:1981,
  	title = {Period Doubling and Chaotic Behavior in a Driven Anharmonic Oscillator},
  	author = {Linsay, Paul S.},
  	journal = {Phys. Rev. Lett.},
  	volume = {47},
  	issue = {19},
  	pages = {1349--1352},
  	numpages = {0},
  	year = {1981},
  	month = {Nov},
  	publisher = {American Physical Society},
  	doi = {10.1103/PhysRevLett.47.1349},
  	url = {https://link.aps.org/doi/10.1103/PhysRevLett.47.1349}
}

@Article{Aharonov:1959,
  	title = {Significance of Electromagnetic Potentials in the Quantum Theory},
  	author = {Aharonov, Y. and Bohm, D.},
  	journal = {Phys. Rev.},
  	volume = {115},
  	issue = {3},
  	pages = {485--491},
  	numpages = {0},
  	year = {1959},
  	month = {Aug},
  	publisher = {American Physical Society},
  	doi = {10.1103/PhysRev.115.485},
  	url = {https://link.aps.org/doi/10.1103/PhysRev.115.485}
}

@Book{Gibbs:1902,
  	title={Elementary principles in statistical mechanics: developed with especial reference to the rational foundations of thermodynamics},
  	author={Gibbs, Josiah Willard},
  	year={1902},
  	publisher={C. Scribner's sons}
}

@Article{Lyubich:2002,
	author="Lyubich, Mikhail",
	title="Almost Every Real Quadratic Map Is Either Regular or Stochastic",
	journal="Ann. Math.",
	year="2002",
	month="2024/11/29/",
	publisher="Annals of Mathematics",
	volume="156",
	number="1",
	pages="1--78",
	issn="0003486X",
	doi="10.2307/3597183",
	url="https://doi.org/10.2307/3597183",
	url="http://www.jstor.org/stable/3597183"
}

@Inbook{Shiryayev:1993,
	author="Shiryayev, A. N.",
	title="New Metric Invariant of Transitive Dynamical Systems and Automorphisms of Lebesgue Spaces",
	bookTitle="Selected Works of A. N. Kolmogorov: Volume III: Information Theory and the Theory of Algorithms",
	year="1993",
	publisher="Springer Netherlands",
	address="Dordrecht",
	pages="57--61",
	isbn="978-94-017-2973-4",
	doi="10.1007/978-94-017-2973-4_5",
	url="https://doi.org/10.1007/978-94-017-2973-4_5"
}

@Article{Kolmogorov:1959,
  	author= {Kolmogorov, A. N.},
  	title = {Entropy per unit time as a metric invariant of automorphism},
  	journal = {Doklady of Russian Academy of Sciences},
  	year = {1959},
  	volume = {124},
  	pages = {754--755}
}

@Article{Sinai:2010,
  	title={On the Notion of Entropy of a Dynamical System},
  	author={Yakov G. Sinai},
  	year={1959},
  	journal   = {Doklady of Russian Academy of Sciences},
  	volume    = {124},
  	pages     = {768--771}
}

@Article{Nambu:1960,
  	title = {Quasi-PArticles and Gauge Invariance in the Theory of Superconductivity},
  	author = {Nambu, Yoichiro},
  	journal = {Phys. Rev.},
  	volume = {117},
  	issue = {3},
  	pages = {648--663},
  	numpages = {0},
  	year = {1960},
  	month = {Feb},
  	publisher = {American Physical Society},
  	doi = {10.1103/PhysRev.117.648},
  	url = {https://link.aps.org/doi/10.1103/PhysRev.117.648}
}

@Article{Goldstone:1961,
	author="Goldstone, J.",
	title="Field theories with Superconductor solutions",
	journal="Il Nuovo Cim. (1955-1965)",
	year="1961",
	month="Jan",
	day="01",
	volume="19",
	number="1",
	pages="154--164",
	issn="1827-6121",
	doi="10.1007/BF02812722",
	url="https://doi.org/10.1007/BF02812722"
}

@Article{Goldstone:1962,
  	title = {Broken Symmetries},
  	author = {Goldstone, Jeffrey and Salam, Abdus and Weinberg, Steven},
  	journal = {Phys. Rev.},
  	volume = {127},
  	issue = {3},
  	pages = {965--970},
  	numpages = {0},
  	year = {1962},
  	month = {Aug},
  	publisher = {American Physical Society},
  	doi = {10.1103/PhysRev.127.965},
  	url = {https://link.aps.org/doi/10.1103/PhysRev.127.965}
}

@article{Simoyi:1982,
  title = {One-Dimensional Dynamics in a Multicomponent Chemical Reaction},
  author = {Simoyi, Reuben H. and Wolf, Alan and Swinney, Harry L.},
  journal = {Phys. Rev. Lett.},
  volume = {49},
  issue = {4},
  pages = {245--248},
  numpages = {0},
  year = {1982},
  month = {Jul},
  publisher = {American Physical Society},
  doi = {10.1103/PhysRevLett.49.245},
  url = {https://link.aps.org/doi/10.1103/PhysRevLett.49.245}
}

@article{Meiss:1992,
  title = {Symplectic maps, variational principles, and transport},
  author = {Meiss, J. D.},
  journal = {Rev. Mod. Phys.},
  volume = {64},
  issue = {3},
  pages = {795--848},
  numpages = {0},
  year = {1992},
  month = {Jul},
  publisher = {American Physical Society},
  doi = {10.1103/RevModPhys.64.795},
  url = {https://link.aps.org/doi/10.1103/RevModPhys.64.795},
  note={and references therein}
}

@book{chernov:2006,
  title={Chaotic billiards},
  author={Chernov, Nikolai and Markarian, Roberto},
  number={127},
  year={2006},
  publisher={American Mathematical Soc.}
}
\end{document}